# Controlled Inertial Cavitation as a Route to High Yield Liquid Phase Exfoliation of Graphene

Piers Turner[1*], Mark Hodnett[1], Robert Dorey[2] and J. David Carey[3,4*]

[1]Ultrasound and Underwater Acoustics, National Physical Laboratory, Teddington, Middlesex, TW11 0LW, United Kingdom.

[2]Centre for Engineering Materials, Department of Mechanical Engineering Sciences, University of Surrey, Guildford, Surrey, GU2 7XH, United Kingdom.

[3]Advanced Technology Institute, University of Surrey, Guildford, Surrey, GU2 7XH, United Kingdom.

[4]Department of Electrical and Electronic Engineering, University of Surrey, Guildford, Surrey, GU2 7XH, United Kingdom.

Corresponding authors

*P.T. email    piers.turner@npl.co.uk

*J.D.C. email    david.carey@surrey.ac.uk

**Abstract:**

Ultrasonication is widely used to exfoliate two dimensional (2D) van der Waals layered materials such as graphene. Its fundamental mechanism, inertial cavitation, is poorly understood and often ignored in ultrasonication strategies resulting in low exfoliation rates, low material yields and wide flake size distributions, making the graphene dispersions produced by ultrasonication less economically viable. Here we report that few-layer graphene yields of up to 18% in three hours without introduction of basal plane defects can be achieved by optimising inertial cavitation during ultrasonication. We demonstrate that the yield and the graphene flake dimensions exhibit a power law relationship with inertial cavitation dose. Furthermore, inertial cavitation is shown to preferentially exfoliate larger graphene flakes which causes the exfoliation rate to decrease as a function of sonication time. This study demonstrates that measurement and control of inertial cavitation is critical in optimising the high yield sonication-assisted aqueous liquid phase exfoliation of size-selected nanomaterials. Future development of this method should lead to the development of high volume flow cell production of 2D van der Waals layered nanomaterials.

Since the discovery of graphene[1] and the characterisation of its properties[1-3], it has shown a huge potential in applications ranging from energy storage[4], solar cells[5], printed electronics[6], composite fillers[7], and, recently, hair dye[8]. The discovery of graphene has also generated significant research interest into other 2D van der Waals layered nanomaterials such as the family of metallic and semiconducting transition metal dichalcogenides[9]. One of the main challenges limiting the further applications and commercialisation of graphene and other 2D layered materials, is that it remains difficult to produce large quantities of high quality flakes with a well-controlled size distribution. Many of the useful properties of graphene are indeed dependent on the lateral size and thickness of individual flakes[3, 5]; for example, graphene flakes with large lateral dimensions ($> 1\ \mu$m) are used in polymer composites[10] and conductive graphene inks[6], flakes with smaller lateral dimensions ($< 1\ \mu$m) are employed in ceramic composites[11], and graphene quantum dots ($< 100$ nm) are found in photovoltaics, fuel cells, and catalysis applications[12].

One of the most scalable dispersed graphene production routes is liquid phase exfoliation from graphite using ultrasonication[13, 14], shear mixing[15], or microfluidization[16]. Such production methods are generally characterised by dispersions with wide (nm-$\mu$m) flake size distributions and low yields of typically 1-5%. As a consequence of the low yields, large unexfoliated graphite flakes are often present in the graphene dispersions post sonication, and extensive centrifugation is required to remove them. Although cascade centrifugation has been shown to be effective in isolating narrow size distributions, it is time intensive, lowers the bulk concentration of the dispersed graphene and can inadvertently remove graphene flakes with larger lateral dimensions[17]. As liquid phase exfoliation techniques typically produce dispersions with low intrinsic graphene concentrations ($\sim 0.1$ mg/ml), centrifugation and re-dispersion is often required to produce graphene dispersions with industrially viable concentrations ($\geq 1$ mg/ml). As such, the removal of large graphene flakes can be unavoidable when producing graphene dispersions.

Despite microfluidization and shear mixing demonstrating superior graphene exfoliation rates[15, 16], ultrasonication is one of the most widely used methods to produce high quality graphene dispersions due to the abundance of sonic baths. The main limitations of using ultrasonication are its wide flake size distributions, low exfoliation rates, and comparatively low graphene yields. This is due to a poor understanding of the fundamental mechanisms driving graphene exfoliation, and a reliance upon purely empirical parameters such as sonication time, temperature calorimetry and nominal electrical input power to monitor and develop ultrasonication strategies. Here, we demonstrate that inertial cavitation is the fundamental mechanism driving graphene exfoliation during ultrasonication. We show that optimisation of inertial cavitation results in dramatically improved graphene yields as high as 18%. We demonstrate the controlled application of inertial cavitation is critical for optimising sonication assisted liquid phase exfoliation of graphene. We further show that inertial cavitation preferentially exfoliates larger flakes during sonication, which results in a saturation of the graphene exfoliation rate as function of sonication time. These findings will be instrumental in developing advanced ultrasonication strategies that will increase the large volume production and commercialisation of a wide range of 2D nanomaterials.

**Acoustic cavitation**

Acoustic cavitation is the stimulated expansion and collapse of microbubbles in response to an applied acoustic field (Fig. 1a). Sonic baths and sonic horns generate acoustic cavitation by exciting a fluid with continuous or pulsed pressure waves at kHz frequencies. A regime with cavitating bubbles that have long lifetimes is referred to as stable cavitation, whereas inertial cavitation is characterised by short lived cavitating bubbles which undergo violent and chaotic collapse[18]. Both types of cavitation exhibit physicochemical effects which are strongly dependent on the properties of the liquid being sonicated (acoustic impedance) as well as the acoustic field frequency, amplitude and geometry[19]. Stable cavitation generates short range vortices known as microstreaming, whereas inertial cavitation collapses radiate spherical shockwaves with velocities of up to 4,000 m/s with peak pressures of up to 6 GPa[20]. Intense liquid jets (jetting) with pressures of up to 1 GPa can also be generated during inertial collapse[21]. In a typical sonication environment, such as a sonic bath or sonotrode, both types of cavitation can exist simultaneously.

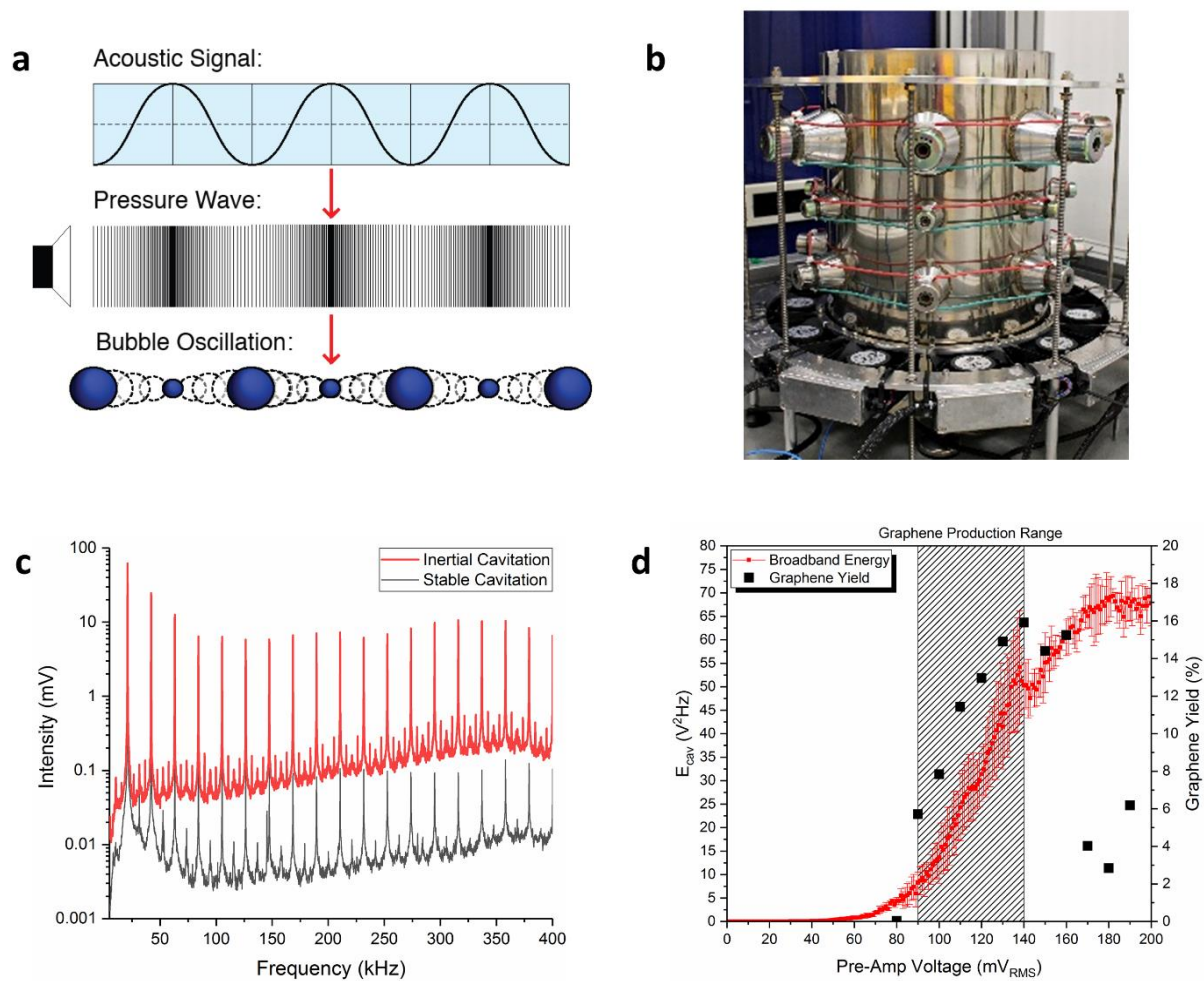

**Fig. 1 | Acoustic cavitation metrology tools and measurements.** (a) Schematic of the growth and collapse of cavitating bubbles in response to an applied acoustic field and the associated pressure wave. (b) Photograph of the National Physical Laboratory's 17 litre multi-frequency reference cavitating vessel with transducers around the circumference. (c) The frequency spectrum of cavitation signals arising from stable and inertial cavitation. The presence of harmonic activity is indicative of stable cavitation activity and the rise in the background noise is indicative of inertial cavitation activity. (d) $E_{cav}$ and the graphene yield as a function of the pre-amp voltage (the output voltage of a signal generator that was used to drive the reference vessels top row of 21.06 kHz transducers via a 400 W power amplifier). The hashed rectangle represents the pre-amp voltage range over which graphene was produced in this study. The uncertainty in $E_{cav}$ is associated with the standard deviation of five independent measurements.

In this work, acoustic cavitation was generated by a multi-frequency reference cavitating vessel (Fig. 1b) that is capable of producing stable and reproducible cavitation fields with a well-defined acoustic field distribution[22]. The acoustic signals that cavitating bubbles emit (Fig. 1c) were measured using a calibrated needle hydrophone. Inertial cavitation was delineated from stable cavitation by quantifying the broadband noise, over a MHz frequency range in which harmonic activity is not distinguishable from the background noise[23]. This was carried out by measuring the high frequency broadband energy[24] (equation 1), parametrised as $E_{cav}$.

$$E_{cav} = \frac{1}{N}\sum_{t=1}^{N} \int_{f_1}^{f_2} V_c(f)^2 df \qquad (1)$$

where $V_c(f)$ are the spectral magnitudes measured from the frequency domain cavitation spectra (Fig. 1c) and $f_1$ and $f_2$ are 1.5 MHz and 2.5 MHz respectively. The inertial cavitation threshold was determined by measuring $E_{cav}$ as a function of the nominal input power of the vessel; the measurement protocol is described in the Supplementary Information 1. As shown in Fig. 1d, the inertial cavitation threshold is characterised by a systematic rise in $E_{cav}$[23]. This occurs above a pre-amp voltage of ~60 mV$_{RMS}$, which corresponds to a nominal input electrical power of 5 Watts (corresponding to a vessel power density of around 0.3 W/L).

To study graphene exfoliation arising from the physiochemical effects of acoustic cavitation, samples were produced by sonicating graphite in 28 ml low density polyethylene vials positioned in a region where inertial cavitation activity is intense and localised (Fig S1.2a). Preliminary experiments found that graphene is first produced only after the onset of the inertial cavitation (Fig. 1d), which demonstrates that the physiochemical effects of inertial cavitation drive graphene exfoliation during ultrasonication. At high pre-amp voltages (high acoustic powers) $E_{cav}$ saturates due to cavitation shielding[25], where a significant volume fraction of cavitating bubbles dynamically scatter and absorb the acoustic field. This considerably affected the graphene exfoliation rate such that a sharp reduction in the graphene yield occurred when $E_{cav}$ saturates (Fig. 1d). The highly non-linear nature of inertial cavitation combined with the significant perturbation of the graphene exfoliation rate at high acoustic powers (Fig. 1d) suggests that measurement and control of inertial cavitation is essential when developing sonication methodologies.

**Inertial cavitation dose**

To evaluate the role of inertial cavitation on the liquid phase exfoliation of graphene, graphite with a narrow 45-75 $\mu$m size distribution (Supplementary Information 2.3), was exfoliated over a range of both pre-amp voltages (shown in Fig. 1c) and sonication times. As $E_{cav}$ is a direct and real time measurement of the inertial cavitation activity, which is the stimulus driving the liquid phase exfoliation of graphene, multiplying $E_{cav}$ by the total sonication time, t, quantifies the accumulated dose of inertial cavitation, (ICD), experienced by the graphite and graphene flakes during sonication. This method for characterising the accumulated inertial cavitation activity has also featured in medical ultrasound studies with analogous measurement protocols[26-28]. As the value of $E_{cav}$, and therefore ICD is dependent on the waveform capture settings (vertical resolution, timebase and sampling rate), the MHz frequency band over which it is calculated, the frequency response of the hydrophone, and the hardware filtering and amplification in the signal chain (Fig. S1.1), absolute values obtained are arbitrary, though the units of ICD can be considered as volts squared. However, as the $E_{cav}$ measurements in this work were carried out using the same measurement protocol, the resultant ICD measurements are directly comparable.

Figure 2a shows that the graphene yield has a power law relationship with ICD, such that there is a linear relationship between the graphene yield and the square root of the ICD (Fig. 2b). As the ICD is a product of $E_{cav}$ and sonication time, this square root relationship explains the observation of graphene yield increasing as a function of the square root of sonication time[13]. The graphene yield, given by $((c_g/c_{gi}) * 100)$, where $c_g$ is the graphene concentration (Supplementary Information 1) and $c_{gi}$ is the graphite concentration, can thus be further enhanced by either increasing the inertial cavitation intensity or the sonication time. The highest graphene yield measured was ~18%, which is amongst the highest in the graphene production literature. During exfoliation trials, it was found that sonication times beyond 120 minutes resulted in anomalously high graphene yields that appeared to deviate from the power law relationship with ICD (Fig. S2.6).

This was subsequently found to occur due to the temperature of the water in the LDPE vials increasing during long periods of sonication (Fig. S2.7), resulting in faster cavitating bubble growth rates and therefore a greater frequency of inertial cavitation collapses[29]. Stable temperatures were maintained over long (150 and 180 minute) sonication times by using an array of cooling fans to actively cool the vessel. As such, cooling strategies should be carefully considered when developing large volume sonication methodologies to ensure consistent yields. Shorter sonication times as well as burst mode ultrasound may also help mitigate temperature increase during sonication.

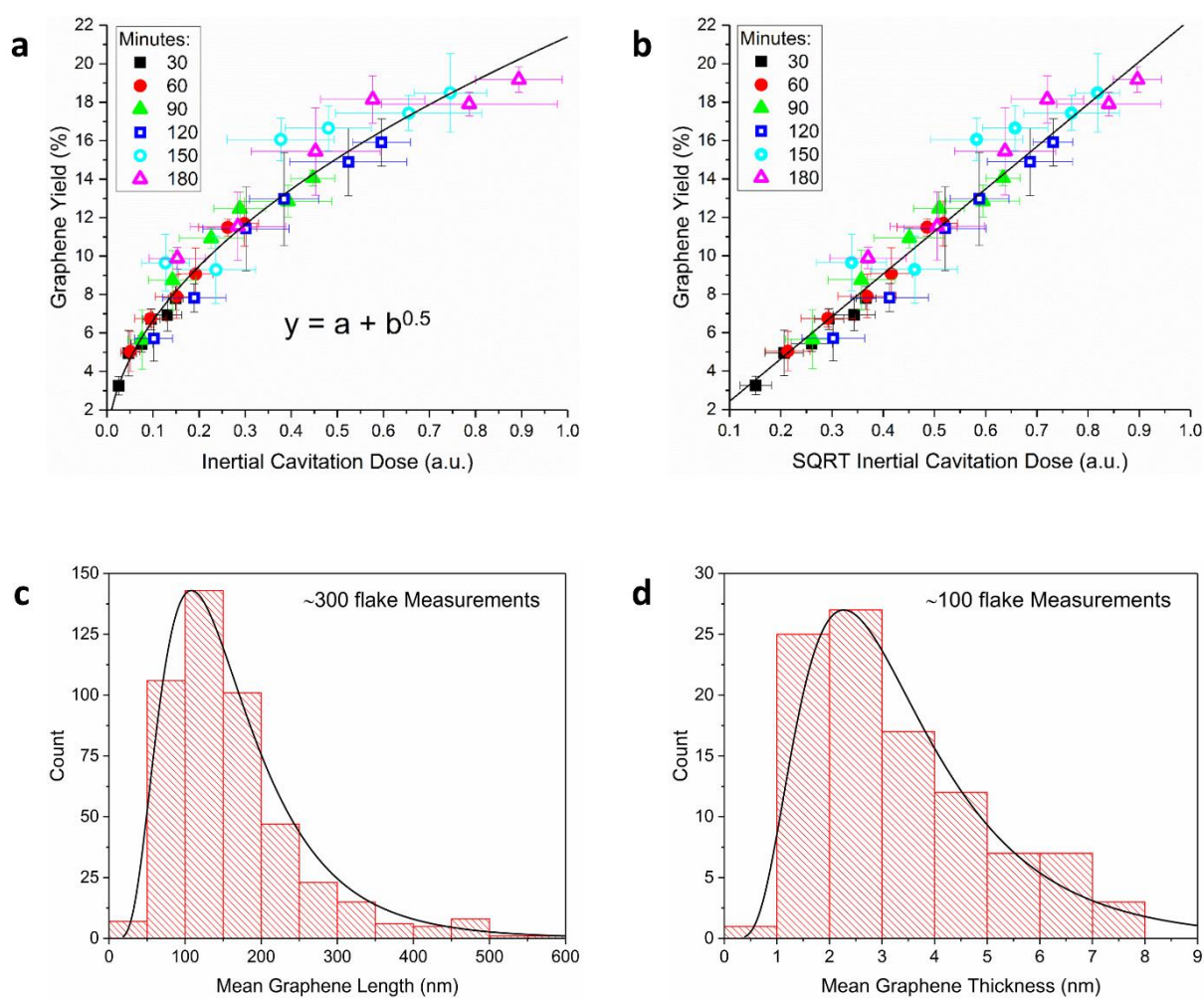

**Fig. 2 | Graphene yield and flake size distribution data.** The graphene yield (after centrifugation) as a function of the (a) ICD and (b) the square root of the ICD. The symbols in (a) and (b) delineate the data as function of sonication time, measured in minutes. Representative lognormal plots of (c) graphene length and (d) thickness distributions that were measured using SEM and AFM, respectively. The uncertainty in the graphene yield is associated with the standard deviation of three graphene yield measurements at each sonication time and the uncertainty in ICD is associated with the standard deviation of five independent high frequency broadband energy measurements.

To quantitatively investigate the evolution of the graphene size distribution as a function of ICD, the lengths and thicknesses of graphene flakes were measured using scanning electron microscopy (SEM) and atomic force microscopy (AFM). The graphene length (Fig. 2c) and thickness (Fig. 2d) distributions were log-normal in shape, implying a multiplicative stochastic fracturing mechanism, whereas a bimodal distribution would be indicative of an erosion process[30]. Accordingly, it can be concluded that inertial cavitation, which is characterised by stochastic and energetic bubble collapse, fractures graphite/graphene during sonication in a stochastic multiplicative process. Figure 3a and 3b show that the mean length and thickness of graphene flakes decreases linearly as a function of the square root of the ICD. As the $< 1 \mu$m lateral dimensions of the flakes in post-sonication precipitate (Fig. S2.5) are significantly smaller than the dimensions of the initial graphite population ($45 - 75 \mu$m), this indicates that the initial graphite population has been fractured by the accumulated inertial cavitation activity during sonication. This finding is consistent with a previous study, which demonstrated that carbon nanotube exfoliation and length reduction was also dependent on inertial cavitation activity[31]. During sonication the mean flake size will progressively decrease until the flakes are small enough to be suspended in the solution by the electrostatic repulsion of the adsorbed sodium cholate surfactant molecules. This is demonstrated in Figs. 3c and 3d, which show that the mean graphene length and thickness are both linearly correlated (Pearson's R ~0.9) with the graphene yield. The linear relationships between the graphene exfoliation rate and the flake size measurements in Fig. 3e and 3f indicate that the physical size of the graphite /graphene flakes limits the rate at which it is exfoliated during ultrasonication. This finding demonstrates that inertial cavitation preferentially exfoliates larger flakes during ultrasonication. Such a size preference, likely arises from the increased size and or surface area of larger flakes, which absorb a greater fraction of the shockwave energy generated by nearby inertially cavitating bubbles. Larger graphite flakes will also have an increased probability of containing structural defects such as holes or tears, resulting in a greater fracturing potential. Furthermore, the length and thickness correlations with the ICD suggest that controlling inertial cavitation may allow for in-situ size control of the graphene size distribution during sonication. With further yield optimisation, the need for centrifugation could also be negated.

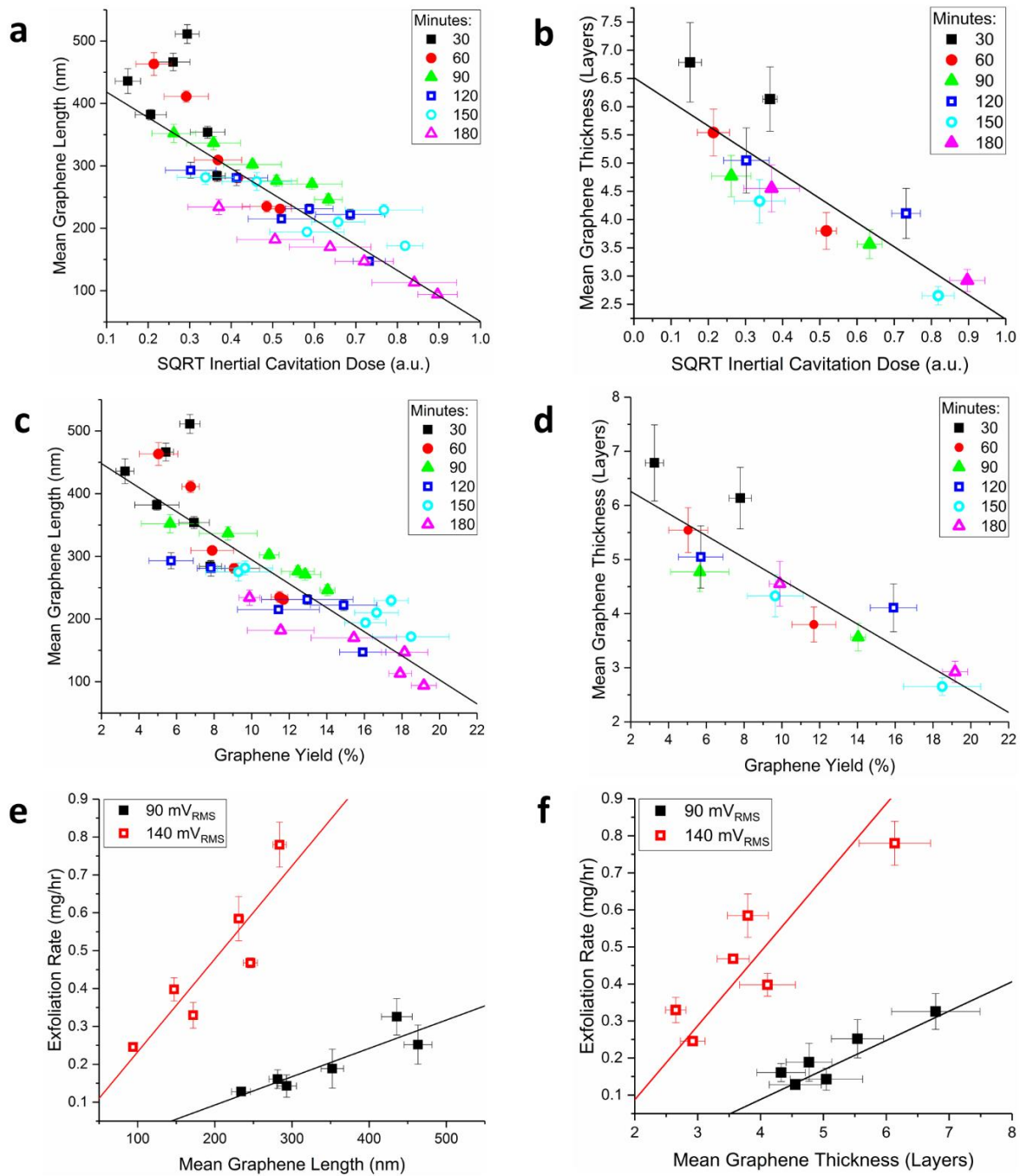

**Fig. 3 | Evidence that inertial cavitation preferentially exfoliated larger flakes during sonication.** The mean graphene (a) length and (b) thickness as function of the ICD. The mean graphene (c) length and (d) thickness as a function of the graphene yield. The symbols in (a-d) delineate the data as function of sonication time, measured in minutes. The graphene exfoliation rate ($c_g/t$) as function of (e) the mean graphene length and (f) thickness for graphene samples produced with the highest and lowest pre-amp voltages (acoustic powers) used in this work. The uncertainty in the graphene yield and exfoliation rate is associated with the standard deviation of three graphene yield measurements at each sonication time, and the uncertainty in graphene length and thickness is associated with the standard error of ~300 and ~100 measurements, respectively.

Graphene exfoliation is likely to be driven by a combination jetting, microstreaming and shockwaves during sonication. As jetting is facilitated by nearby extended surfaces, and the maximum size of the graphite used (sieved to 45-75 $\mu$m) is much smaller than the resonant size of the cavitating bubbles found in this work (~160 $\mu$m at 21.06 kHz), jetting events within the graphene dispersion will significantly decrease in frequency as the mean flake size decreases during sonication. However, jetting will continue to exfoliate any graphite/graphene flakes on the inner walls of the LDPE vials. The local shear stresses generated by collapsing bubbles, known as microstreaming, are also unlikely to drive graphene exfoliation as the speed of microstreaming vortices is proportional to the squared frequency of cavitating bubbles[32], making microstreaming more effective at megasonic frequencies[33]. Consequently, the shockwaves generated by inertial cavitation are the most probable exfoliation mechanism during sonication. Shockwave exfoliation will be mediated by a combination fracturing events triggered by incident shockwaves[36] and the high velocity interparticle collisions that are generated by incident shockwaves[37, 38]. However, as shockwaves lose more than 50% of their initial energy over the first 25 $\mu$m of propagation due to absorption[20], and will be attenuated by dispersed graphene flakes (which will increase in density during sonication), graphene exfoliation is most likely facilitated by the shockwaves generated by immediately adjacent inertial cavitation collapse events.

**Raman characterisation for material quality and flake size quantification**

Spectroscopic trends in the graphene samples were explored using Raman spectroscopy. To ensure that the Raman spectra were representative of the graphene dispersions, 120 spectra were collected and averaged across 20 x 20 $\mu$m areas of re-stacked graphene films. These films were produced by filtering graphene dispersions through alumina membranes (20 nm pore size) using vacuum filtration, and contain dense ordered networks (~50% free volume) of nanosheets. As such, the laser beam (532 nm excitation wavelength) interrogates 100s of flakes per Raman measurement[39]. The Raman spectrum of graphene, shown in Fig. 4a, consists of the characteristic G, D, and 2D peaks, which are indicative of many properties including functionalisation, strain, defects, and size distribution[39-43]. Specifically it has been established that the ratio of the intensity of the D peak (~1320 cm$^{-1}$) to that of the G peak (~1580 cm$^{-1}$) is indicative of graphene nanosheet size, and the shape and intensity of the 2D peak (~2700 cm$^{-1}$) changes as function of graphene nanosheet thickness[39, 42, 44].

Figure 4a shows that the ratio of the G peak to the D peak intensity, $I_D/I_G$, increases between the minima and maxima of the ICD range. As the inverse of the $I_D/I_G$ ratio is proportional to the length of graphene flakes[39], this shows that the graphene length is decreasing over the ICD range. This is further quantified in Fig. 4b, which shows that the $I_D/I_G$ ratio increases as a function of the square root of the ICD. Although this trend mirrors the quantitative SEM length distributions in Fig. 3a, it is less strongly correlated. This is likely due to the intentionally low centrifugation speeds used in this work (1000 rpm, 120g), that were chosen to minimise the effects of centrifugation on the dispersed graphene population. The resulting wide size distributions are evidenced by the large variances in $I_D/I_G$ that were observed across each two-dimensional Raman map (Fig. S3.2a). This is likely the cause of the large uncertainties and anomalies in the $I_D/I_G$ ratio (Fig. 4b) measured from graphene samples produced at a high ICD. The $I_{2D}/I_G$ intensity ratio was ~0.45 across the entire ICD range, suggesting a mean graphene thickness of ~5 layers[39]. However the $I_{2D}/I_G$ ratio also had large variances across each two-dimensional map (Fig. S3.2b). This finding is inconsistent with the quantitative AFM size distribution data shown in Fig. 3b, and suggests that Raman measurements made on restacked graphene films can yield unreliable flake thickness and length metrics when the graphene dispersions contain larger flakes and/or wide size distributions. However, beyond $I_D/I_G$ and $I_{2D}/I_G$, Backes *et al.*[17] demonstrated that other spectroscopic metrics such as the full width at half maximum of the G peak (Fig. 4c) and the ratio of the 2D peak to the graphite peak shoulder ($I_{2720}/I_{2690}$, Fig. S3.3) are also indicative of nanosheet size. Figure 4c shows that the full width at half maximum of the G peak, $\Gamma_G$, increases as a function of the square root of the ICD. As $\Gamma_G$ was found to be inversely proportional to nanosheet size[17], this finding is consistent with the SEM data shown in Fig. 3a.

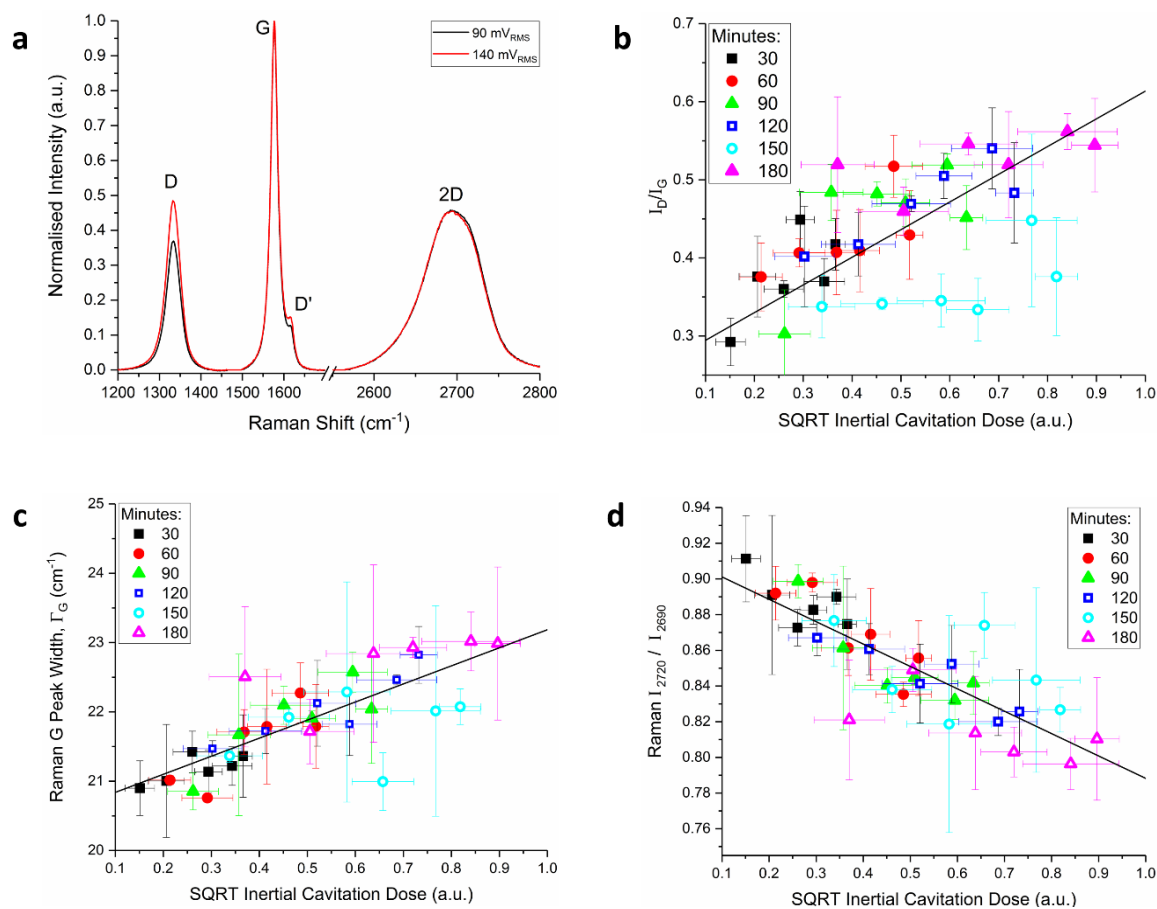

**Fig. 4 | Raman spectra and spectroscopic size distribution metrics.** (a) Normalised Raman spectra of graphene dispersions produced at the maxima and minima of the ICD range. (b) The $I_D/I_G$ ratio as a function of the square root of the ICD. (c) The full width at half maximum of the Raman G Peak as function of the square root of the ICD. (d) The ratio of the graphite 2D peak (~2720 cm$^{-1}$) to the graphite 2D peak shoulder (~2690 cm$^{-1}$) as a function of the square root of the ICD. The uncertainty in the $I_D/I_G$ ratio, $\Gamma_G$, $I_{2720}/I_{2960}$, and the ICD is associated with the standard deviation of three Raman measurements and five broadband energy measurements per sample, respectively.

Furthermore, Fig. 4d shows that the $I_{2720}/I_{2690}$ decreases as a function of the square root of the ICD. As $I_{2720}/I_{2690}$ is proportional to the graphene thickness, this is also consistent with the AFM data shown in Fig. 3b. Alongside graphene size distribution analysis, Raman spectroscopy can also be used to quantify the defects in graphene. Eckmann *et al.*[43] reported that the ratio of the D peak to the D' peak was dependant on the defect type, such that an $I_D/I_{D'}$ ratio of ~13, ~7, and ~3 indicates the presence of sp$^3$ defects, vacancy defects and, edge defects respectively. As $I_D/I_{D'}$ was found to be ~3 for all graphene samples in this work, this shows that inertial cavitation causes flake scission without introducing a significant number of basal plane defects in graphene during sonication.

**Conclusion**

By optimising the inertial cavitation dose graphene yields of up to 18% have been produced in just three hours. These high graphene yields, which are produced by ultrasonication, were achieved over relatively short sonication times with minimal temperature increases and low nominal input powers. We demonstrate that the graphene yield as well as the graphene flake length and thickness exhibits a power law relationship with inertial cavitation dose, which is a direct measurement of the violent collapses that are indicative of inertial cavitation. During sonication graphite is fractured in a multiplicative process by the shockwaves generated by immediately adjacent inertial cavitation activity; few basal plane defects are incorporated. We also show that temperature increase during sonication can result in inconsistent exfoliation, and therefore temperature control strategies should be employed to ensure consistent ultrasonication. More generally, we show that careful measurement and control of acoustic cavitation is critical when developing efficient ultrasonication methodologies and which by extension can ultimately lead to high volume flow cell production of 2D van der Waals layered nanomaterials.

**Methods**

Fine Flake Graphite purchased from Asbury Carbons was pre-treated by sonicating in a 1 litre LDPE container for 30 minutes in an Ultrawave IND1750 sonic bath at a concentration of 10 mg/ml, with 1 litre of 15 MΩ de-ionised water and 1 mg/ml of sodium cholate surfactant (Sigma Aldrich). The graphite was then vacuum dried and sieved through a 75 $\mu$m test sieve to remove large flakes and then sieved using a 45 $\mu$m test sieve to remove small flakes, resulting in a size distribution of 45-75 $\mu$m. 0.2 mg/ml of pre-treated and sieved graphite was added to a 28 ml LDPE Nalgene vial (Fisher Scientific) along with a 25 ml magnetically stirred solution of de-ionised water with 3 mg/ml of sodium cholate surfactant (Sigma Aldrich). The LDPE vials were pre-soaked in a water and surfactant solution (0.2% vol of Cole Palmer MICRO-90) overnight prior to sonication to promote wetting of the external surface of the vials. The reference vessel was actively cooled using an array of 12 V fans when sonicating graphene samples over long durations. To facilitate cavitation, surfactant (Cole Palmer MICRO-90) was added at a 0.2% by-volume concentration to the bulk volume of the vessel. After sonication, the graphene dispersions were left to sediment overnight before being centrifuged at 1000 rpm (120 rcf) for 2 hours. The supernatant was then removed and characterised. The acoustic cavitation measurement details, experimental methodology development, and the graphene characterisation methods are described in detail in the supplementary information.

**Acknowledgments**

This work was supported by the U.K. National Measurement System (NMS) under the innovation R&D programme and by the U.K. Engineering and Physical Sciences Research Council (EPSRC) via the Centre for Doctoral Training in Micro and Nano Materials and Technology at the University of Surrey. The authors would like to thank Dr. B. Zeqiri, Dr. A. Pollard, Dr. K. Paton, Dr A. Wain, Dr. A. Sesis and Ms. E. Legge for useful discussions and technical assistance in sample preparation, characterisation and data analysis.

**Author contributions**

P.T. carried out the graphene production and analysis work. M.H., R.D. and J.D.C. supervised the project. M.H. advised on the use of NPL's multi-frequency reference cavitation vessel and assisted in interpreting acoustic cavitation measurement data. All authors contributed to the interpretation of the results and the writing of the manuscript.

**Data Access**

Details of the data and how to request access are available from the University of Surrey Publications Repository.

**Competing financial interest**

Aspects of the work presented here have been included on a patent application GB1812056.8.